\documentclass[a4paper,fleqn,usenatbib]{mnras}

\usepackage{mathptmx}
\usepackage{txfonts}

\usepackage[T1]{fontenc}
\usepackage{ae,aecompl}

\usepackage{graphicx}	

\newcommand*{\hnull}{\ensuremath{H_{0}}}
\newcommand*{\lnull}{\ensuremath{\lambda_{0}}}
\newcommand*{\onull}{\ensuremath{\Omega_{0}}}
\newcommand*{\eg}{e.g.}
\newcommand*{\etal}{et~al.}
\newcommand*{\ie}{i.e.}
\newcommand*{\perse}{per se}
\newcommand*{\ack}{ACKNOWLEDGEMENTS} 
\newcommand*{\bestfit}{best-fitting}
\newcommand*{\pmzd}{parametrized} 
\newcommand*{\Fig}{Fig.}
\newcommand*{\Figs}{Figs.}
\newcommand*{\Tab}{Table}
\newcommand*{\fig}{fig.}
\newcommand*{\sect}{section}
\newcommand*{\cdash}{-} 
\newcommand*{\kdash}{ -- } 
\newcommand*{\pdash}{~--~} 
\newcommand*{\jdash}{--} 
\newcommand*{\rdash}{--} 
\newcommand*{\Universe}{Universe} 
\newcommand*{\percent}{per cent}
\newcommand*{\software}[1]{{\sc #1}}

\title[Supernovae and scatter in distance moduli]%
{The $m$\jdash$z$ relation for Type~Ia supernovae: safety in numbers or
safely without worry?} 

\author[P.~Helbig]{Phillip Helbig\thanks{E-mail: 
helbig@astro.multivax.de}
\\
Thomas-Mann-Str.~9, D-63477 Maintal, Germany}

\date{Accepted 2015 August 3.  Received 2015 July 27; 
in original form 2015 June 30}

\pubyear{2015}

\begin{document}
\label{firstpage}
\pagerange{\pageref{firstpage}\rdash\pageref{lastpage}}
\maketitle

\begin{abstract}
The $m$\jdash$z$ relation for Type~Ia supernovae is compatible with the
cosmological concordance model if one assumes that the \Universe\ is
homogeneous, at least with respect to light propagation.  This could be
due to the density along each line of sight being equal to the overall
cosmological density, or to `safety in numbers', with variation in the
density along all lines of sight averaging out if the sample is large
enough.  Statistical correlations (or lack thereof) between redshifts,
residuals (differences between the observed distance moduli and those
calculated from the \bestfit\ cosmological model), and observational
uncertainties suggest that the former scenario is the better
description, so that one can use the traditional formula for the
luminosity distance safely without worry. 
\end{abstract}



\section{Introduction}

I recently investigated the dependence of constraints on the
cosmological parameters \lnull\ and \onull\ derived from the
$m$\jdash$z$ relation for Type~Ia supernovae on the degree of local
homogeneity of the \Universe\ 
\citep{PHelbig15Ra}.  
When deriving such constraints, it is often assumed that the \Universe\ is
completely homogeneous, at least with regard to light propagation.
However, the constraints on the cosmological parameters derived depend
on this assumption.  If the degree of local inhomgeneity is \pmzd\ by
the parameter $\eta$ giving the fraction of homogeneously distributed
matter on the scale of the beam size such that the density at a given
redshift is equal to the average cosmological density
$\rho=\frac{3H^{2}\Omega}{8\upi{}G}$ outside the beam and $\eta\rho$
inside the beam 
\citep*[see][for definitions and discussion]{RKayserHS97Ra}, 
and assuming that $\eta$ is independent of redshift and the same for all
lines of sight, then only if $\eta\approx{}1$ do the constraints on the
cosmological parameters \lnull\ and \onull\ derived from the
$m$\jdash$z$ relation for Type~Ia supernova correspond to the
`concordance model' 
\citep[\eg][]{JOstrikerPSteinhardt95a,EKomatsuetal2011a,PLANCKXVI2014a}.

Two important conclusions of 
\citet{PHelbig15Ra}
are thus that the values of \lnull\ and \onull\ derived from the
$m$\jdash$z$ relation for Type~Ia supernovae depend on assumptions made
about $\eta$, substantially so for current data, and that only for
$\eta\approx{}1$ are these values consistent with other measurements of
the cosmological parameters. 
\citet{SPerlmutteretal99a}
considered the effect of $\eta\neq1$ on their results (see their \fig~8
and the discussion in their \sect~4.3) and concluded that, at least in
the `interesting' region of the \lnull--\onull\ parameter space
(\ie~$\onull<1$; even at that time there was substantial evidence
against $\onull>1$), it had a negligible effect.  Not only is this
effect no longer negligible with newer data (both because there are more
data points altogether and because there are more data points at higher
redshifts), but, especially since we now have good estimates of \lnull\
and \onull\ from other tests, it allows one to use the supernova data to
say something about $\eta$.  With a strong indication from the supernova
data that $\eta\approx{}1$, it is important to consider the question
whether this is true only when several lines of sight are averaged or is
true for a typical individual line of sight. 
\citet{SPerlmutteretal99a} 
investigated the influence of $\eta$ on the values obtained for \lnull\
and \onull\ but could draw no conclusions about its value from the
supernova data alone.  Even though the `concordance model' had already
been postulated at the time (though of course there was much less
evidence in favour of it than is the case today), assuming the
corresponding values for \lnull\ and \onull\ could not allow any
statement to be made about $\eta$ since there was significant overlap in
the allowed regions of parameter space for the various $\eta$ scenarios.
(Also, in contrast to the case with newer data, the \bestfit\ values of
\lnull\ and \onull\ were far from the concordance values, though the
concordance values were allowed even at 1~$\sigma$.)  This is consistent
with their claim, based on simulations, that the conclusions drawn from
their data should not depend heavily on $\eta$.  Their robust conclusion
that the $m$\jdash$z$ relation for Type~Ia supernovae implies that
$\lnull>0$, and the somewhat stronger claim that $q_{0}<0$ (\ie~the
\Universe\ is currently accelerating), regardless of assumptions made
about $\eta$, are of course the most interesting results of 
\citet{SPerlmutteretal99a}
(and similar studies by the High\rdash$z$ Supernova Search Team and
later papers by both groups).  Interestingly, both of these are still
robust with current data.

\section{Two scenarios}

There are two ways in which $\eta\approx{}1$ can be explained.  One is
that $\eta\approx{}1$ holds for each individual line of sight.  (This
does not necessarily imply that $\eta$ is actually constant along the
beam, but only that the distance modulus calculated from the
cosmological parameters \lnull, \onull, and \hnull\ and from the
redshift $z$ is the same as that calculated assuming $\eta\approx{}1$.
In other words, there could be density variations along the beam (apart
from the decrease in density with decreasing redshift due to the
expansion of the \Universe) as long as they appropriately average out.)
The other is that $\eta<1$ for some lines of sight and $\eta>1$ for
others, such that $\eta\approx{}1$ when averaged over all lines of
sight, though of course density variations along the beam as in the
other case could also be present.\footnote{The second case requires a
more general definition of $\eta$ than that used in 
\citet{RKayserHS97Ra}; 
see 
\citet*{JLimaVS14a} 
and 
\citet{PHelbig15Ra} 
for discussion.  Strictly speaking, as pointed out by 
\citet{SWeinberg76a},
it is the magnification $\mu$ which averages to 1 over all lines of
sight.  Since $\eta\sim\kappa$, where $\kappa$ is the convergence, and
$\mu\sim(1-\kappa)^{-1}$, the relation is linear only in the limit of
vanishing deviations, though approximately linear for the small
deviations considered here.  The actual situation is quite complicated.
For example, the average angular-size distance $\langle{}D\rangle$, and
hence the average luminosity distance $\langle{}D_{\mathrm{L}}\rangle$,
is biased even in the case of $\langle\mu\rangle=1$.  See 
\citet{NKaiserJPeacock2015a}
for discussion of this and many other details in the still ongoing
debate on this topic.  I use the term `average' here loosely; the
important point is that the average of an observed quantity is the same
as in the $\eta=1$ case, not that $\eta$ itself averages to 1.} This has
been dubbed the `safety in numbers' effect by 
\citet{DHolzELinder05a}.  

In the first case, the residuals (the differences between the observed
distance moduli and those calculated from the \bestfit\ cosmological
parameters) should not depend on redshift \perse, while in the second
case they should increase with redshift: all else being equal, the lower
$\eta$, the larger the distance modulus, and the difference between this
and that calculated using the traditional $\eta=1$ assumption is a
monotonically increasing function of redshift; see \eg\ \fig~1 in 
\citet{RKayserHS97Ra}.\footnote{There is of course a similar effect with 
opposite sign for $\eta>1$, as discussed in the previous footnote.} In
the first case, the residuals are due only to uncertainties in the
observed distance moduli, while in the second case they are due also to
variations in the actual distance moduli as a result of different
average densities along the line of sight.  Of course, in the first case
there could be a dependence of the residuals on redshift if the
observational uncertainties depend on redshift, and in the second case
the residuals are due both to variations in the actual distance moduli
and to observational uncertainties in them.  One could call the first
case `safely without worry', meaning that one can safely use the
traditional formula for the luminosity distance (corresponding to
$\eta=1$) when calculating the distance modulus, without worry.

\section{Calculations, results, and discussion}

For purposes of comparison and consistency, I work with the same data as
in 
\citet{PHelbig15Ra}, 
namely the publicly available `Union2.1' sample of supernova data 
\citep{NSuzukietal2012a}. 
\begin{figure}
\includegraphics[width=0.8\columnwidth]{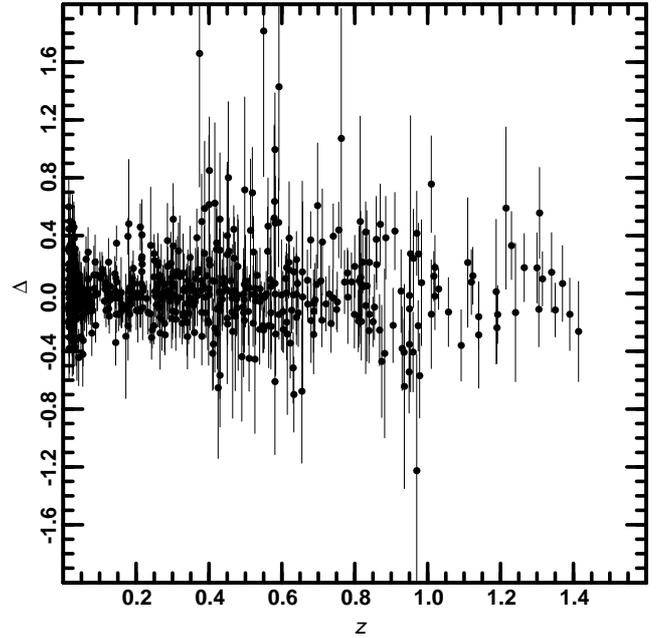}
\caption{Residuals (differences between the observed distance moduli and
those calculated from the \bestfit\ cosmological model) (points) and
observational uncertainties (lines).} 
\label{residuals_with_errors}
\end{figure}
\Fig~\ref{residuals_with_errors} shows the residuals $\Delta$ (points)
with respect to the \bestfit\ model assuming $\eta=1$ in 
\citet{PHelbig15Ra}
($\lnull=0.7210938$ and $\onull=0.2773438$) and the uncertainties
$\sigma$ in the distance moduli (lines).  These are shown separately
(both as points) in 
\begin{figure}
\includegraphics[width=0.8\columnwidth]{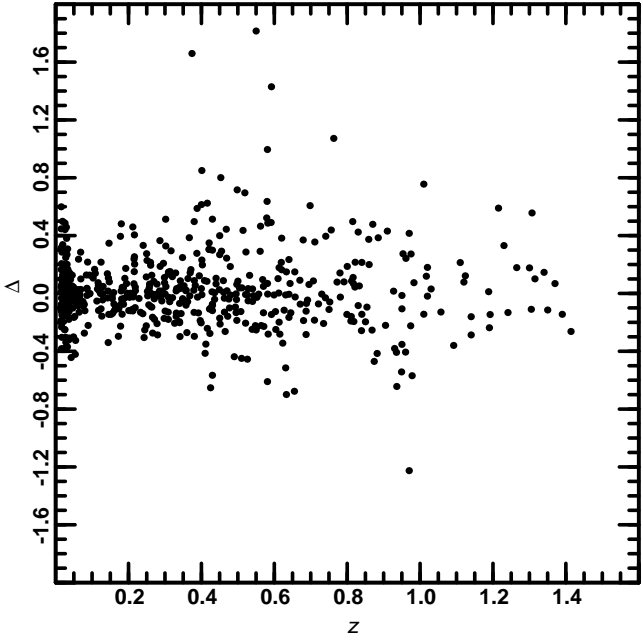}
\caption{Residuals.}
\label{residuals}
\end{figure}
\Figs~\ref{residuals} and 
\begin{figure}
\includegraphics[width=0.8\columnwidth]{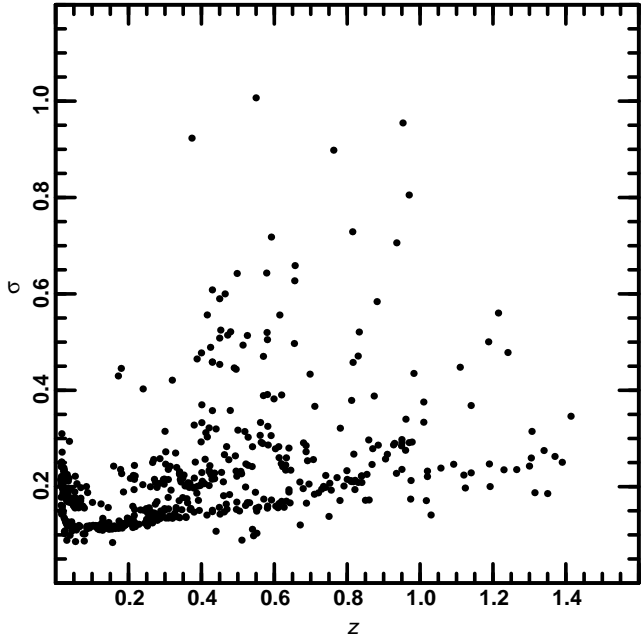}
\caption{Observational uncertainties.}
\label{errors}
\end{figure}
\ref{errors}.  There appear to be a positive correlation between the
uncertainties and redshifts and a higher number of outliers at
intermediate redshifts (though the fact that there are fewer at high
redshifts might be due to the smaller number of objects there).  If the
first case discussed above holds, then (the absolute value of) the
quotient $Q=\Delta/\sigma$ of the residuals and the uncertainties should
show no trend with redshift, while if the second case holds there should
be a positive correlation. 
\begin{figure}
\includegraphics[width=0.8\columnwidth]{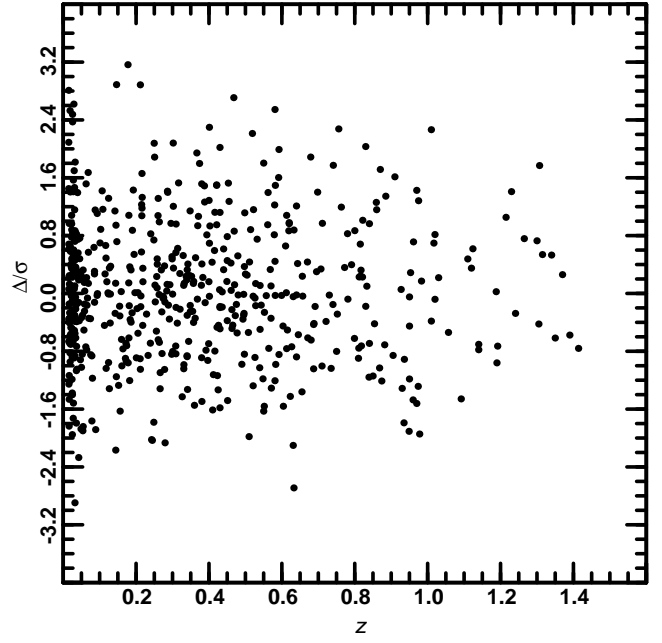}
\caption{Quotients of residuals and observational uncertainties.}
\label{scaled_residuals}
\end{figure}
\Fig~\ref{scaled_residuals} shows this quotient and, indeed, there
appears to be no trend with redshift.  Also, the width of the
distribution seems to depend only on the number of points in the
corresponding redshift range, \ie\ there appear to be no outliers as
such, or at least fewer. 

In order to quantify the dependence of the magnitude of the
uncertainties on redshift, I have calculated various statistical
measures, shown in \Tab~\ref{table}, to investigate the existence of a
correlation between the redshifts and the absolute values of the
residuals $|\Delta|$ 
\begin{figure}
\includegraphics[width=0.8\columnwidth]{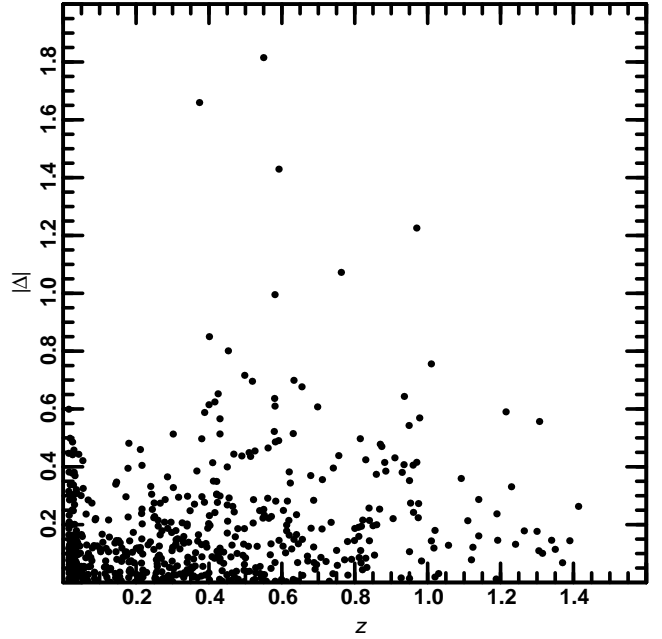}
\caption{Absolute values of residuals.}
\label{absolute_residuals}
\end{figure}
(plotted in \Fig~\ref{absolute_residuals}), the observational
uncertainties $\sigma$ (\Fig~\ref{errors}), and the absolute value of
the quotient of the residuals and the uncertainties, $|Q|$
\begin{figure}
\includegraphics[width=0.8\columnwidth]{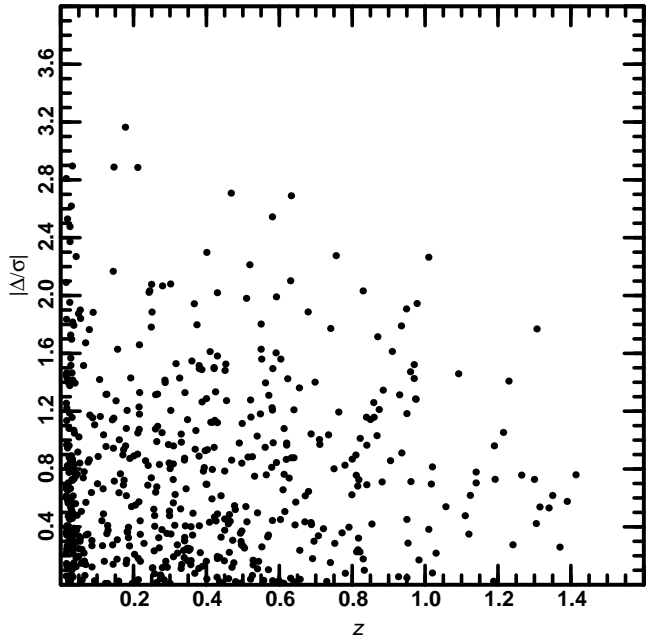}
\caption{Quotients of absolute values of residuals and observational
uncertainties.} 
\label{absolute_scaled_residuals}
\end{figure}
(plotted in \Fig~\ref{absolute_scaled_residuals}), as well as the
corresponding statistical significance.  Note that
\Fig~\ref{absolute_residuals}, like \Fig~\ref{errors}, appears to show a
positive correlation between the absolute values of the residuals and
redshifts and a higher number of outliers at intermediate redshifts. 
Also included in \Tab~\ref{table} are the corresponding quantities
concerning the correlation between $|\Delta|$ and $\sigma$. 
\begin{table*}
\centering
\caption{Statistical quantities measuring the correlation between the
redshifts $z$ and the absolute values of the residuals $|\Delta|$, the
observational uncertainties $\sigma$, and the quotient $|Q|$ of these,
as well as between $|\Delta|$ and $\sigma$: $r$ is Pearson's
product\cdash{}moment correlation coefficient, $r_{\mathrm{s}}$ is
Spearman's rank\cdash{}order correlation coefficient, and $\tau$ is
Kendall's non\cdash{}parametric rank\cdash{}order correlation
coefficient.  The corresponding $p$ values give the probability of
getting a value as large as observed or larger in the case of the null
hypothesis of no correlation.  All values have been rounded to two
significant figures.} 
\label{table}
\begin{tabular}{lllllll}
\hline
data set & $r$ & $p(r)$ & $r_{\mathrm{s}}$& $p(r_{\mathrm{s}})$ & $\tau$ &
$p(\tau)$ \\ 
\hline
$z,|\Delta|$    
         & 0.23 & $1.3\times10^{-8}$
         & 0.21 & $1.8\times10^{-7}$
         & 0.14 & $7.8\times10^{-7}$ \\
$z,\sigma$
         & 0.41 & $1.5\times10^{-25}$
         & 0.44 & $4.5\times10^{-29}$
         & 0.28 & $2.6\times10^{-23}$ \\
$z,|Q|$
         & $3.6\times10^{-2}$ & 0.39
         & $5.1\times10^{-2}$ & 0.22
         & $3.3\times10^{-2}$ & 0.24 \\
$|\Delta|,\sigma$
         & 0.62 & 0.00
         & 0.42 & $1.3\times10^{-25}$
         & 0.29 & $1.7\times10^{-25}$ \\
\hline
\end{tabular}
\end{table*}

All three statistical tests agree about the sign of the correlation and
whether or not it is significant.  (The \emph{values} of the
correlations and the corresponding significance are not directly
comparable.)  Both the absolute values of the residuals, $|\Delta|$, and
the observational uncertainties, $\sigma$, are positively correlated
with redshift, but their quotient is not.  This suggests that the first
scenario described above, `safely without worry', is the appropriate
one, not the second scenario, `safety in numbers'.  If this is the case,
then one would expect $|\Delta|$ and $\sigma$ to be correlated, and
indeed they are.  (Note that this last test is not sufficient to rule
out the `safety in numbers' scenario, since even if the scatter in the
actual distance moduli increased with redshift, there could still be a
correlation between $|\Delta|$ and $\sigma$ in addition to the one
between $|\Delta|$ and $z$.) 

Of course, this analysis takes the Union2.1 data set at face value, and
relies on the assumption that the observational uncertainties have been
correctly estimated.  Also, $\eta\neq{}1$ describes just a different
amount of Ricci focusing due to more or less matter within the beam
than in the standard case, as opposed to more general gravitational
lensing.  Note that 
\citet{NSuzukietal2012a} 
explicitly correct for the amplification of supernovae known to be
gravitationally lensed by galaxy clusters (see their \sect~2.1); in
other words, the magnitudes used for the cosmology analysis are those
which would have been observed in the absence of the corresponding
galaxy clusters.  To be sure, 
\citet{NSuzukietal2012a}, 
following the procedure described in \sect~7.3.5 of
\citet{RAmanullahetal2010a},
include as part of the error estimate $0.093z$ to take the statistical
uncertainty due to gravitational lensing into account.  If this were a
significant part of the uncertainty, then it could explain the
correlation between the uncertainties and redshifts, and thus favour the
`safety in numbers' scenario.  This contribution to the error budget
probably explains the slope of the lower envelope in \Fig~\ref{errors}.
However, it is clear from \Figs~\ref{errors} and \ref{scaled_residuals}
that the main cause of the correlation is the absence of both large
residuals and large uncertainties at low redshifts.  The large
residuals\pdash{}much larger than $0.093z$\pdash{}also have large
uncertainties, and occur mainly at intermediate redshifts.  Finally, as
described in \sect~7.2 of 
\citet{RAmanullahetal2010a} 
and \sect~4.4 of
\citet{NSuzukietal2012a}, 
the Union2.1 data set was constructed by rejecting 3-$\sigma$ outliers,
which would remove any strongly lensed supernovae from the sample.  Both
this use of median statistics and the $0.093z$ contribution contribute
to the correlation at some level but, as explained above, cannot explain
all, or even most, of it. 

Note that 
\citet{Yuetal11a}, 
using observational data other than the $m$\jdash$z$ relation for
Type~Ia supernovae, and assuming a flat \Universe, arrive at essentially
the same conclusion as 
\citet{PHelbig15Ra}: 
$\eta\approx{}1$ is favoured and low values of $\eta$ can be ruled
out.\footnote{%
\citet{Yuetal11a} 
usually refer to (Ruth)
\citet{RDalyetal08a} 
as `Ruth \etal'.}  Some of the assumptions in 
\citet{Yuetal11a} 
were questioned by 
\citet{VBustiRSantos11a}, 
but even when these are corrected for, $\eta\approx{}1$ is still
favoured.  As discussed in 
\citet{PHelbig15Ra}, 
one expects to measure a larger value of $\eta$ when larger angular
scales, such as those investigated in 
\citet{Yuetal11a}, 
are considered, so the result of 
\citet{PHelbig15Ra} 
remains interesting because of the small angular scales of supernova
beams. 

The 
\citet{PLANCKXV2015a}
measured the CMB lensing-deflection power spectrum at 40\,$\sigma$,
showing it to agree with the smooth $\Lambda$CDM amplitude (\ie~the
$\eta=1$ case) to within 2.5~\percent.  Since all forms of gravitating
clumps contribute to this, such a measurement of the power as a function
of scale is fairly definitive about the smoothness of the energy-density
distribution.  This should be contrasted with the situation a few
decades ago, when it was widely believed that there was no dark matter
other than that required for flat rotation curves in spiral galaxies and
for bound galaxy clusters; $\eta\approx{}0$ was thought to be the best
approximation even for objects as large as large galaxies 
\citep[\eg][]{JGottGST74a,RRoeder75a}.
To be sure, most of the analysis done by the 
\citet{PLANCKXV2015a}
deals with $L\leq400$, although $L<2048$ is also investigated, where $L$
is the multipole.  $L=400$ corresponds to an angular scale of somewhat
less than a degree and $L=2048$ to about 10\,arcmin.  This means that
the corresponding physical size in the concordance model is about 5\,Mpc
at $z=1$ (and about 40\,kpc at the redshift of the CMB).  Thus, the
$m$\jdash$z$ relation for Type~Ia supernovae probes much smaller scales,
and indicates that even at these scales $\eta\approx{}1$ is appropriate,
\ie\ that the \Universe\ is homogeneous at even these very small scales.

\section{Conclusions}

There is a statistically significant correlation between the absolute
value of the residuals, \ie\ the difference between the observed
distance moduli and those calculated from the \bestfit\ cosmological
model, and the observational uncertainties in the Union2.1 sample of
Type~Ia supernova observations.  Each of these quantities is also
correlated with redshift but their quotients are not.  This suggests
that each individual line of sight to these supernovae is a fair sample
of the \Universe\ in the sense that the (average) density is approximately
the same as the overall density; in other words, it is not necessary to
average over several lines of sight in order to recover the overall
density.  Since most of the matter in the \Universe\ is dark matter, it
must be distributed smoothly enough so that most lines of sight contain
the same density as the overall average density. When the resolution of
cosmological numerical simulations becomes high enough to resolve the
corresponding scale, this distribution must result.  Rather than putting
it in `by hand', it would be more interesting if it emerged from other
assumptions or theoretical considerations.

\section*{\ack}

I thank an anonymous referee for helpful comments.  Figures were
produced with the \software{gral} software package written by Rainer
Kayser.

\bibliographystyle{mnras}
\bibliography{etasnia2}

\bsp	
\label{lastpage}
\end{document}